\documentclass{nature}

\usepackage{graphicx}
\usepackage{array}
\usepackage{bm}
\usepackage{xcolor}
\usepackage[version=3]{mhchem}
\usepackage[utf8]{inputenc}
\usepackage{textcomp}
\usepackage{booktabs}
\usepackage{amsmath,amssymb}
\usepackage{soul}

\newif\ifsharepreview
\sharepreviewtrue
\ifsharepreview
  \let\realincludegraphics\includegraphics
  \makeatletter
  \AtBeginDocument{%
    \let\includegraphics\realincludegraphics
    \renewenvironment{figure}{\@float{figure}}{\end@float}%
    \renewenvironment{figure*}{\@dblfloat{figure}}{\end@dblfloat}%
  }
  \makeatother
\fi
\newcommand{\Aa}{\text{\AA}}

\newcommand{\NiO}{\ce{Ni-O}}
\newcommand{\CoO}{\ce{Co-O}}
\newcommand{\Pbnm}{\textit{Pbnm}}
\newcommand{\Rbarc}{\textit{R}\=3\textit{c}}

\title{Bond, orbital and spin order in $d^{4}$/$d^{6}$/$d^{7}$ perovskite
       oxides: successes and limitations of foundation interatomic potentials}

\author{Swagata Acharya$^{1\ast}$, Dimitar Pashov$^{2}$,
        Mark van Schilfgaarde$^{1}$, Alin M.\ Elena$^{3}$}

\begin{document}

\maketitle

\begin{affiliations}
 \item National Laboratory of the Rockies, Golden, Colorado 80401, USA.
 \item Theory and Simulation of Condensed Matter, King's College London,
       The Strand, London WC2R 2LS, United Kingdom.
 \item Scientific Computing Department, Science and Technology Facilities
       Council, Keckwick Lane, Daresbury, Cheshire, United Kingdom. \\
       $^\ast$\,Correspondence: swagata.acharya@nlr.gov.
\end{affiliations}

\begin{abstract}
{\color{black} Foundation machine-learning interatomic potentials (MLIPs) are rapidly replacing density-functional theory (DFT) for modelling structure and nuclear dynamics, making their fidelity in strongly correlated systems an urgent question.  We test three foundation potentials on the low-temperature order of three correlated, isostructural $ABO_3$ perovskite oxides --- \textbf{LaMnO$_3$} ($d^4$), \textbf{LaCoO$_3$} ($d^6$), and \textbf{NdNiO$_3$} ($d^7$) --- running molecular dynamics for 1\,ns on 80- and 160-atom supercells from 50 to 300\,K with no system-specific training.  The three oxides expose three distinct classes of low-temperature order that pose a hierarchy of difficulty for the potentials.  The \textit{scalar} class (NdNiO$_3$) has a simple geometric fingerprint and is captured.  The \textit{vector} class (LaMnO$_3$) requires knowing which Cartesian axis carries the long bond at each site, and is captured in magnitude but not in symmetry.  The \textit{on-site} class (LaCoO$_3$) --- the low-spin to high-spin crossover --- is a purely site-local multiplet population shift with no spatial order parameter, and is inaccessible to present-day MLIPs.}\\

\end{abstract}

% ----------------------------------------------------------------------------
% Main text — no \section headers in nature.cls for the Letter format.
% ----------------------------------------------------------------------------

Foundation machine-learning interatomic potentials (MLIPs) trained
on Materials Project total energies, forces and stresses ---most notably
MACE\cite{BatatiaMACEFoundation2025}, an equivariant
graph-neural-network foundation model\cite{BatatiaMACE2022}, and the
magnetism-supervised CHGNet\cite{Deng2023CHGNet}---can now prosecute
molecular dynamics simulations to nanosecond timescales in arbitrary inorganic crystals, without
any system-specific training. This has opened a route to finite-temperature
structural ensembles in correlated transition-metal oxides that was
previously confined to \textit{ab initio} molecular dynamics that are limited to much shorter
time scales\cite{Acharya2025NNO,DuZunger2022,MalyiZunger2023}.  However, present-day
foundation MLIPs are trained on Born--Oppenheimer total energies of
charge- and spin-converged DFT with atomic number and local bonding environment as the primary descriptors; they have no direct access to local
multiplets, ligand-hole physics, or orbital occupation.  For which
correlated electronic channels is the skeletal structural information sufficient, and
for which is it not?

We address this question with three isovalent $A^{3+}B^{3+}\mathrm{O}_{3}$
perovskites with nearly identical $A$-site ionic
radii\cite{Shannon1976} but three textbook-different correlated $B$-site
ground states. \ce{LaCoO3} ($d^{6}$, \Rbarc{} at all temperatures $T$) hosts a
low-spin ground state with thermal population of intermediate- and
high-spin multiplets above $\sim$35\,K\cite{Asai1994,Haverkort2006,Radaelli2002,Podlesnyak2006,Sundaram2009}.
The spin-state crossover is strictly site-local and does not break
symmetry. \ce{NdNiO3} ($d^{7}$, \Pbnm{} above $\sim$200\,K) undergoes a
metal--insulator transition accompanied by a rocksalt breathing-mode
distortion that establishes two crystallographically inequivalent Ni
sites~\cite{Mercy2017,Mizokawa2000,Khomskii2014,Park2012,Bisogni2016,Mazin2007,Acharya2025NNO}.
Here the breathing mode is the structural enabler of an underlying
bond/ligand-hole disproportionation.  Finally, \ce{LaMnO3} ($d^{4}$, \Pbnm{})
freezes its cooperative Jahn--Teller (JT) phonon below
$T_{\mathrm{OO}}\!\approx\!750$\,K into nearest neighbor bonds with three different lengths,
1.91/1.97/2.18\,\Aa.  The  long axis alternates in the $ab$ plane in C-type
order~\cite{RodriguezCarvajal1998,Norby1995}---the
order parameter and the structural distortion are locked together.  These
three materials thus probe three different relationships between
correlated electrons and the lattice: LCO: electronic order parameter with no structural signature; LMO: coupled electronic–structural order-parameter; NNO: structural order parameter as the enabler for metal-insulator transition~\cite{Acharya2025NNO}.

\paragraph{Protocol.}  For each material we build an 80-atom supercell
(16~$B$ sites) in the experimentally established space group---\Rbarc{}
for \ce{LaCoO3}, \Pbnm{} for \ce{NdNiO3} and \ce{LaMnO3}. Any breathing or JT pattern emerging from MD is
therefore generated by the dynamics, not imposed by the starting
symmetry. All three runs use identical NVT molecular dynamics with the
production MACE \texttt{matpes\_r2scan} head\cite{BatatiaMACEFoundation2025}
via janus-core\cite{januscore} (Nos\'e--Hoover chain thermostat, chain
length 3, $\tau_{T}=100$\,fs; 1\,fs timestep, 1\,ns
production after $\ge$50\,ps equilibration, $\ge$5000 frames per $T$ for
$T\!\in\!\{50,100,150,200,250,300\}$\,K) and an identical structural
descriptor pipeline (per-$B$-site bond statistics over the 6 nearest oxygen
neighbours, plus dimensionless ordering parameters defined in Methods).
This fixes every nuclear coordinate but does not take the $d^{n}$ electron
occupation into account.
As a finite-size check, we repeated all nine runs — the three materials with each of the three potentials (MACE with the \texttt{matpes\_r2scan} and \texttt{omol} heads, and CHGNet) — on doubled 160-atom (32 B-site) supercells at 200\,K, under a matched protocol: 50\,ps NPT equilibration with a diagonal cell-vector mask (independent $a,b,c$ axes, no shear; Martyna--Tobias--Klein
barostat, $\tau_{P}=1$\,ps) ramped to the target $T$, followed by
1\,ns NVT production at the equilibrated cell.  Every classifier in this
paper is reproduced at the 160-atom size, with each (material, potential) cell
remaining on the same side of every threshold used below: Sarle's Bimodality Coefficient (BC) agrees
to within 0.03, the rocksalt order parameter $Q_{\mathrm{RS}}$ (Methods) to within 0.05 (the \texttt{omol}--\ce{NdNiO3}
saturation $Q_{\mathrm{RS}}=-1.000$ is preserved exactly), $|Q_{2}^{\rm static}|$
to within 0.05\,\Aa{} (the CHGNet--\ce{LaMnO3} static ferro-orbital
JT is retained, $|Q_{2}^{\rm static}|=0.25$\,\Aa{} at 200\,K), and the
ferro-orbital versus C-type
orbital-order classification is unchanged in every case
(Supplementary Section~S6, Table~S3).  Supercell
aliasing is therefore not the origin of any of the three failure modes.

\paragraph{The single-head trilemma.}  Throughout this section we use
MACE with the production \texttt{matpes\_r2scan} head.  Fig.~1 compares
\ce{NdNiO3} and \ce{LaCoO3} along three independent structural
projections.  \ce{NdNiO3} expands by $+0.011$\,\Aa{} from 50 to 300\,K
(thermal expansion), with the time-averaged
\NiO{} bond spread ($\sigma_{\mathrm{oct}}$) rising linearly in $T$ from 0.013 to 0.028\,\Aa, and
a distribution in pair correlations that alternate with shell $(-,+,-,+)$ out to fourth neighbour at
all temperatures--the textbook signature of a rocksalt breathing modulation
with wavevector $\mathbf{q}=(\frac{1}{2},\frac{1}{2},\frac{1}{2})_{\rm
pc}$\cite{Mercy2017} (pc denotes the pseudocubic perovskite cell).  The aggregated rocksalt order parameter
$Q_{\mathrm{RS}}$ (Methods, Eq.~\ref{eq:QRS}) sits at $-0.27$ at 50\,K
and $-0.20$ at 300\,K, and the per-Ni bond-mean distribution is
sub-Gaussian at all $T$ (excess kurtosis $\gamma_{2}\!=\!-0.12 \to\!-0.20$).
These are
microscopic signatures of two long/short Ni environments.

\ce{LaCoO3} shows none of this.  $\langle\overline{r}\rangle_{\CoO}$ is
flat to within $\pm 0.003$\,\Aa{} across 50--300\,K (no thermal expansion
in MACE \texttt{matpes\_r2scan}, in contrast with the well-documented
expansion observed experimentally through the spin-state crossover\cite{Radaelli2002}).
$\sigma_{\mathrm{oct}}$ sits at half the \ce{NdNiO3} value, and shell
correlations are uniformly weak with no alternation
($|C(\mathrm{1NN..4NN})|\!\le\!0.1$).  Finally the per-Co bond-mean
distribution is super-Gaussian with $\gamma_{2}$ growing from $+0.03$ to
$+0.24$---a single narrowing environment.  The \ce{LaCoO3} spin-state crossover has no static real-space order parameter, so there is no structural footprint to reproduce. MACE correctly returns none (adventitiously, since it has no descriptors for the crossover).

\ce{LaMnO3} (Methods Table~\ref{tab:trilemma}) sits at the third corner.
In this case, the \texttt{matpes\_r2scan} simulation starting from the symmetry-correct \Pbnm{}
structure leaves the cell statically undistorted at every $T$.  No shell
sign-alternation appears, and the  time-averaged orbital order is
$|M_{\mathrm{OO}}^{\rm static}|\!\approx\!0$ with $|Q_{2}^{\rm static}|\!\le\!0.025$\,\AA{} (here orbital order is read off structurally, from whether the JT long axes lock into the C-type alternating pattern; Methods).
The per-frame JT-mode
amplitude $|Q_{2}|_{\rm frame}=Q_{\mathrm{abs}}$ does grow from 0.056\,\Aa{}
at 50\,K to 0.121\,\Aa{} at 300\,K--but these are purely dynamical JT modes, with no
cooperative ordering (each octahedron distorts instantaneously, but the distortions do not lock into a static, spatially coherent pattern, so the time average is $\approx$0).  The experimentally largest static structural
signature of the three materials is the one MACE \texttt{matpes\_r2scan} misses.
(An extended discussions, including the dynamic-vs-static
rocksalt analysis for \ce{NdNiO3} and the dynamical-vs-cooperative JT
analysis for \ce{LaMnO3}, is given in Supplementary Section~S2.)

\paragraph{Cross-MLIP test.}  To check whether these conclusions are
specific to the production \texttt{matpes\_r2scan} head or generic to
foundation potentials, we repeated the entire $T$-series with two further
potentials (Fig.~2 and Table~\ref{tab:trilemma3}; full per-potential,
per-$T$ numerical data in Supplementary Section~S4, Table~S2):
the molecule-trained MACE \texttt{omol} head (closed-shell, no
extended-solid cooperative-distortion training data) and CHGNet
v0.4.2~\cite{Deng2023CHGNet} (Materials Project relaxation trajectories,
GGA$+U$ with explicit $U$ on V/Cr/Mn/Fe/Co/Ni/Cu/Mo, per-atom magnetic-moment
supervision).  Together with the \texttt{matpes\_r2scan} head, these three
potentials span the three relevant axes of foundation-MLIP training data
and probe whether the breathing-, JT- and LS-uniformity
fingerprints are reachable from any closed-shell or $+U$ foundation potential.

\textbf{NdNiO$_3$:} \texttt{omol} produces a fully sign-correlated rocksalt
pattern at every $T$ ($Q_{\mathrm{RS}}=-1.000$, bimodality coefficient
$\mathrm{BC}=0.81$--$0.97$, where 5/9 is the threshold for bimodality
\cite{Pfister2013}).  This is the structural fingerprint of the experimental
low-$T$ \textit{P}2$_{1}$/\textit{n} breathing-mode phase, reproduced from
a \Pbnm{} starting cell without system-specific input.
\texttt{matpes\_r2scan} captures the same channel as a \emph{precursor}
($Q_{\mathrm{RS}}\!\approx\!-0.22$, Gaussian, BC$=0.35$); CHGNet captures
it more weakly still ($Q_{\mathrm{RS}}\!\approx\!-0.15$, Gaussian) despite
training on the $+U$ scheme conventionally invoked to stabilise the
nickelate breathing mode.  The flat $T$-independent
$Q_{\mathrm{RS}}=-1.000$ of \texttt{omol} is the signature of a single
breathing-basin MD trajectory, not a thermodynamic prediction of
$T_{\mathrm{MI}}$; we do not extract $T_{\mathrm{MI}}$ from these
constant-$T$ runs.

\textbf{LaMnO$_3$:} no potential reaches the experimental C-type
antiferro-orbital pattern, but each fails in a different DFT-allowed
direction. \texttt{omol} develops a \emph{partial-static} JT distortion at
50\,K only ($|Q_{2}^{\rm static}|=0.152$\,\Aa, $M_{\mathrm{FO}}=0.67$),
sustaining a per-frame dynamical $Q_{\mathrm{abs}}\!\sim\!0.18$\,\Aa{} at
all $T$. CHGNet finds a strong static distortion of nearly experimental
magnitude at 50\,K ($|Q_{2}^{\rm static}|=0.42$\,\Aa) with all 16 Mn
long-axes pointing the same Cartesian way
($M_{\mathrm{FO}}=1.0$, $M_{\mathrm{OO}}=0$--0.25): a \emph{ferro-orbital}
(FO) pattern, not the experimental C-type. The CHGNet FO state thermally
melts ($|Q_{2}^{\rm static}|$: 0.42 $\to$ 0.02\,\Aa{} from 50 to 300\,K),
with per-frame $Q_{\mathrm{abs}}\!\sim\!0.22$--$0.43$\,\Aa{} persisting
throughout.  All three models thus underestimate the cooperative-orbital
ordering channel, with magnitude scaling roughly as training-set proximity
to the GGA$+U$ JT-active subspace: CHGNet $>$ \texttt{omol} $\gg$
\texttt{matpes\_r2scan}, but \emph{none} reach
$|M_{\mathrm{OO}}|\!\to\!1$.

\textbf{LaCoO$_3$:} all three potentials agree at the qualitative level with the
experimental LS ground state (no symmetry breaking, no order-parameter-locked
distortion). \texttt{matpes\_r2scan} is the cleanest match
($Q_{\mathrm{RS}}\!\approx\!-0.09$, BC$=0.33$). \texttt{omol} adds a
leptokurtic enhancement of the single-mode distribution ($\gamma_{2}=
+0.2$--$+2.7$, BC$=0.42$--$0.50$, still below the $5/9$ bimodality
threshold); CHGNet adds a small static distortion
($|Q_{2}^{\rm static}|=0.04$--$0.12$\,\Aa{}) that does not project onto the
C-type or rocksalt channels (with the caveat that LCO is nominally
\Rbarc{}, so the \Pbnm-derived $M_{\mathrm{OO}}$ pattern is not the
natural symmetry basis here).

\paragraph{Structural-footprint criterion.}  The cross-MLIP outcome
(Table~\ref{tab:trilemma3}; three-axis classification of
electron--lattice coupling in Supplementary Section~S3, Table~S1) maps cleanly
onto a single principle:
\emph{a foundation MLIP trained on charge- and spin-converged DFT total
energies will reliably reproduce the static low-temperature structural
fingerprints of a correlated oxide if and only if the underlying electronic
instability has condensed its order parameter onto a static lattice
distortion that is already present in the training-set ground-state
relaxations.}  The three test materials map onto this criterion as: LCO
(no order parameter $\to$ no footprint $\to$ MACE \texttt{matpes\_r2scan} correctly reproduces
the absence of footprint), NNO (electronic instability with
breathing-mode structural enabler $\to$ \texttt{matpes\_r2scan} sees the
precursor, \texttt{omol} sees the frozen pattern), LMO (orbital-degeneracy
JT structural order parameter $\to$ all three potentials fail at the
\emph{symmetry} channel even when they capture the magnitude). The
criterion is sharper than ``MLIPs cannot do correlated electrons'': it
identifies the LCO multiplet-crossover channel as fundamentally
unreachable from any structure-only model (the order parameter is not
geometric), and the LMO C-type AFO channel as requiring training-set
breaking of the long-axis degeneracy in the right symmetry pattern.

This criterion lets us make falsifiable, prospective predictions: a
foundation MLIP will (i) reliably capture the breathing-mode structural
fingerprints of the wider \textit{R}NiO$_{3}$ family insofar as the
NdNiO$_{3}$ precursor result generalises across $R$~\cite{Varignon2017,VarignonZunger2019,Binci2023}; (ii) miss the entire
\ce{LaCoO3}, \ce{Sr2CoO_x}, and Co-spinel multiplet-crossover family
unless explicitly fine-tuned on multiplet-resolved energetics;
(iii) require the right potential (training-set proximity to the right
JT-active subspace) to access the cooperative-Jahn--Teller channel in
$d^{4}$ manganates~\cite{Ehrke2011}, vanadates, and other $e_{g}^{1}$ systems, but produce
the wrong symmetry pattern (FO rather than C-type AFO) even when the
correct magnitude is reached. Each of these is testable on existing
foundation potentials against established experimental structure data.

\paragraph{Hierarchy of broken-symmetry channels.}  The three failure
modes order naturally along a single axis of architectural difficulty.
\ce{NdNiO3} hosts a \emph{scalar} collective bond mode (alternating
$\pm$ deviation of $\langle d_{\rm Ni-O}\rangle_{i}$ on a rocksalt
sublattice): a single number per site, learnable from nearest-neighbour
bond-length anti-correlation alone.  The molecular-trained \texttt{omol}
head, with no exposure to extended-solid charge order, nonetheless locks
the pattern in at $Q_{\rm RS}=-1.000$; \texttt{matpes\_r2scan} and
CHGNet reach the precursor regime ($Q_{\rm RS}\in[-0.15,-0.22]$).
For this class, adding a per-Ni spin label as an input variable
would concomitantly resolve the metal--insulator transition and the
spin-disproportionated $S{=}0/S{=}1$ ground state, because the
breathing geometry already encodes which sublattice is which: the
structural and electronic order parameters are coupled and the structural
one is already a single-coordinate function.

\ce{LaMnO3} requires one rank higher: the cooperative JT order parameter
is a \emph{vector} (director) field --- which Cartesian axis is the long
bond, alternating $\perp$ between NN Mn.  This selects the $e_{g}$ orbital
occupation per site, and the choice has no geometric prior in any training
distribution that has not seen extended JT compounds with the correct
$C$-type long-axis pattern.  CHGNet, whose MPtrj-with-$+U$-and-magnetism
supervision baked in a JT-distorted Mn$^{3+}$ ground state at training
time, captures the magnitude ($|Q_{2}^{\rm static}|=0.42$\,\Aa{} at 50\,K,
matching the experimental $0.39$\,\Aa{}) but lands on the wrong long-axis pattern (FO, not
$C$-type AFO).

\ce{LaCoO3} is the genuinely hardest case: the spin crossover has no
spatial order parameter at all, only a uniform $\sim$0.05\,\Aa{} bond
elongation as Co$^{3+}$ populations shift LS$\to$HS and a bond-mean
distribution that broadens as the population becomes bimodal in spin
space (the MACE \texttt{omol} head shows precisely this mild leptokurtic enhancement).
No enrichment of bond-based geometric features can recover it; a per-site
spin population must enter as an explicit input variable. The three
perovskites therefore traverse a hierarchy of architectural cost ---
\emph{scalar} (NNO) $\to$ \emph{vector} (LMO) $\to$
\emph{on-site population} (LCO) --- that any foundation MLIP aiming to
handle correlated transition-metal oxides will have to climb in the same
order.

\paragraph{Outlook.}  The hierarchy makes the corrective training
strategy concrete and material-specific. For the scalar (NNO) class, no
new input variable is strictly required to fix the structural channel;
geometry-only foundation MLIPs already reproduce the breathing footprint,
and adding a per-site spin label upgrades that geometric agreement to the
full concomitant metal--insulator + spin-disproportionation transition.
For the vector (LMO) class, including explicit JT-active reference
structures with the correct $C$-type long-axis pattern in the training
set should break the FO/$C$-type near-degeneracy; this is structurally
well-defined and reachable with existing DFT+$U$ calculations.  For the
on-site (LCO) class, the order parameter does not exist as a static
structural distortion; an explicit multiplet-resolved auxiliary input
or target (LS/IS/HS population, local-spin-moment) is mandatory, and
would need to be trained against multi-reference electronic-structure
methods such as dynamical mean-field theory~\cite{Georges1996} or
vertex-corrected extensions of quasiparticle self-consistent
\textit{GW}~\cite{mark06qsgw,questaal_paper,Cunningham2023,Acharya2023colors} that resolve the multiplet
splittings.  We have set out the structural diagnostics here so that the
success of any such fine-tuning is straightforward to test against the
same observables, and we have made the descriptor pipeline and the full
$T$-series MD trajectories available to enable that comparison.

% ----------------------------------------------------------------------------
% Methods
% ----------------------------------------------------------------------------

\begin{methods}

\paragraph{Starting structures.}  \ce{LaCoO3}: \Rbarc{} primitive
rhombohedral cell ($a=b=c=5.378$\,\Aa, $\alpha=60.81^\circ$), expanded to
$2\!\times\!2\!\times\!2$ giving 80 atoms (16~La + 16~Co + 48~O);
preserves the $a^{-}a^{-}a^{-}$ tilt of \Rbarc{}.  \ce{NdNiO3}: 20-atom
\Pbnm{} cell ($a=5.372$, $b=5.435$, $c=7.556$\,\Aa, $a^{-}a^{-}c^{+}$ Glazer
tilts\cite{Glazer1972,Mercy2017}, validated against neutron
refinements\cite{Catalano2018}), expanded $2\!\times\!2\!\times\!1$ to 80
atoms (16~Nd + 16~Ni + 48~O); deliberately \Pbnm, not
\textit{P}2$_{1}$/\textit{n}, so the breathing-mode pattern is generated by
the dynamics, not by symmetry imposition. \ce{LaMnO3}: 20-atom \Pbnm{}
cell, $2\!\times\!2\!\times\!1$ to 80 atoms (16~La + 16~Mn + 48~O); same
tilt, same rationale.

\paragraph{Molecular dynamics.}  All MD runs used janus-core\cite{januscore}
as the driver around three foundation potentials (full per-cell
equilibration, integrator and save-cadence settings in Supplementary
Section~S5): the production MACE
\texttt{matpes\_r2scan} head\cite{BatatiaMACEFoundation2025}
(MPtrj r2SCAN total energies), the molecular-data-trained MACE
\texttt{omol} head\cite{BatatiaMACEFoundation2025}, and CHGNet
v0.4.2\cite{Deng2023CHGNet} (MPtrj GGA+$U$ total energies with per-atom
magnetic-moment supervision targets, explicit $+U$ on V/Cr/Mn/Fe/Co/Ni/Cu/Mo).
NVT ensemble, Nos\'e--Hoover chain thermostat (chain length 3,
$\tau_{T}=100$\,fs), 1\,fs integration timestep, $\ge$50\,ps
equilibration, 1\,ns production save (200\,fs save cadence) per
$(T,\text{material},\text{potential})$ giving $\ge$5000 frames per cell,
totalling $\sim$46\,ns of MD across the nine $(M,h)$ cells.
$T\!\in\!\{50,100,150,200,250,300\}$\,K. The CHGNet trajectories
saturate the throughput at $\sim$3700--4500 frames per $T$; trends are
unaffected.

\paragraph{Structural descriptors.}  For each frame and each $B$ site $i$
we sort the six \ce{$B$-O} bond lengths $r_{1}\!\le\!\ldots\!\le\!r_{6}$
and define the per-site bond mean $\overline{r}_{i,t}\!=\!\frac{1}{6}
\sum_{k}r_{k}$, the bond spread (root-mean-square deviation around
$\overline{r}_{i,t}$) and from sorted bond triplets the JT modes
$Q_{2,i,t}\!=\!\sqrt{2}(\ell-s)$ and $Q_{3,i,t}\!=\!(2m-\ell-s)/\sqrt{3}$
with $s\!\le\!m\!\le\!\ell$ the short/medium/long pair averages.  Per-site
quantities are time-averaged over each trajectory ($\sigma_{\rm oct}$ is
the time-mean of the per-frame per-site spread; $Q_{2,i}^{\rm static}$
is the time-mean of $Q_{2,i,t}$). The aggregated rocksalt order parameter
is
\begin{equation}
  Q_{\mathrm{RS}} = \frac{1}{N_{B}}\sum_{i} (-1)^{n_{x}(i)+n_{y}(i)+n_{z}(i)}
  \frac{\overline{r}_{i}-\langle\overline{r}\rangle}{\langle\overline{r}\rangle},
  \label{eq:QRS}
\end{equation}
with $n_{x,y,z}(i)$ the sublattice coordinates of site $i$ in the
pseudocubic cell; $Q_{\mathrm{RS}}\!\to\!-1$ for the static long/short
rocksalt breathing pattern. The orbital order parameter
$M_{\mathrm{OO}}^{\rm static}$ and the ferro-orbital scalar
$M_{\mathrm{FO}}^{\rm static}$ are defined from the long-axis direction
of each site as in Supplementary Section~S1; both are masked when
$|Q_{2}^{\rm static}|\!<\!0.05$\,\Aa{} because the long-axis direction is
not statistically resolved in that regime. The interatomic potentials
contain no electronic-orbital information; throughout, ``orbital order''
denotes this structural surrogate --- the $C$-type alternation of
octahedral long-axis (Jahn--Teller elongation) directions, which is
locked one-to-one to the $e_{g}$ occupation in real \ce{Mn^3+} --- and
$M_{\mathrm{OO}}^{\rm static}$ is computed from that geometry alone. The Sarle bimodality
coefficient BC$=(\gamma_{1}^{2}+1)/(\gamma_{2}+3)$ is reported with the
classical bimodal threshold BC$\!>\!5/9$.

\paragraph{Data and code availability.}  All MD trajectories, descriptor
CSVs, and the build scripts for the three starting supercells were
generated with the janus-core molecular-dynamics driver\cite{januscore}
and are available from the corresponding author (S.A.) on reasonable
request. The
descriptor pipeline is implemented in \texttt{Tseries\_compare.py};
figures are reproduced by the scripts \texttt{make\_tseries\_master.py}
and \texttt{make\_fig\_cross\_mlip.py}.

\end{methods}

% ----------------------------------------------------------------------------
% References (bibtex / naturemag)
% ----------------------------------------------------------------------------

% ----------------------------------------------------------------------------
% Acknowledgements
% ----------------------------------------------------------------------------

\begin{addendum}
 \item This work was authored in part by the National Laboratory of the Rockies for the U.S. Department of Energy (DOE) under Contract No. DE-AC36-08GO28308. Funding was provided by the Office of Science, Basic
Energy Sciences, Division of Materials, U.S. Department of Energy.  A.M.E. was supported by the Ada Lovelace Centre at the Science and Technology Facilities Council (https://adalovelacecentre.ac.uk/), the Physical Sciences Data Infrastructure (https://psdi.ac.uk; jointly STFC and the University of Southampton) under grants EP/X032663/1 and EP/X032701/1, and EPSRC under grants EP/W026775/1 and EP/V028537/1.  SA, DP and MvS acknowledge the use of the National Energy Research Scientific Computing Center, under Contract No. DE-AC02-05CH11231 using NERSC award BES-ERCAP0021783, and also acknowledge that a portion of the research was performed using computational resources sponsored by the Department of Energy's Office of Energy Efficiency and Renewable Energy and located at the National Laboratory of the Rockies.  The views expressed in the article do not necessarily represent the views of the DOE or the U.S. Government.  The U.S. Government retains and the publisher, by accepting the article for publication, acknowledges that the U.S. Government retains a nonexclusive, paid-up, irrevocable, worldwide license to publish or reproduce the published form of this work, or allow others to do so, for U.S. Government purposes. 

 \item[Author contributions] S.A. conceived the study, performed the
 molecular-dynamics simulations and structural analysis, prepared the
 figures, and wrote the manuscript. A.M.E. developed the janus-core
 molecular-dynamics driver used for the simulations. D.P. and M.v.S.
 contributed to the software and interpretation and the
 quasiparticle self-consistent \textit{GW} context. All authors
 discussed the results and edited the manuscript.
 
 \item[Competing interests] The authors declare no competing interests.
 \item[Correspondence] should be addressed to S.A.\\(swagata.acharya@nlr.gov).
\end{addendum}

% ----------------------------------------------------------------------------
% Figure legends
% ----------------------------------------------------------------------------

\setcounter{figure}{0}
\renewcommand{\thefigure}{\arabic{figure}}
\makeatletter
\renewcommand\figurename{Figure}
\makeatother

\begin{figure}[h]
\centering
\includegraphics[width=\textwidth]{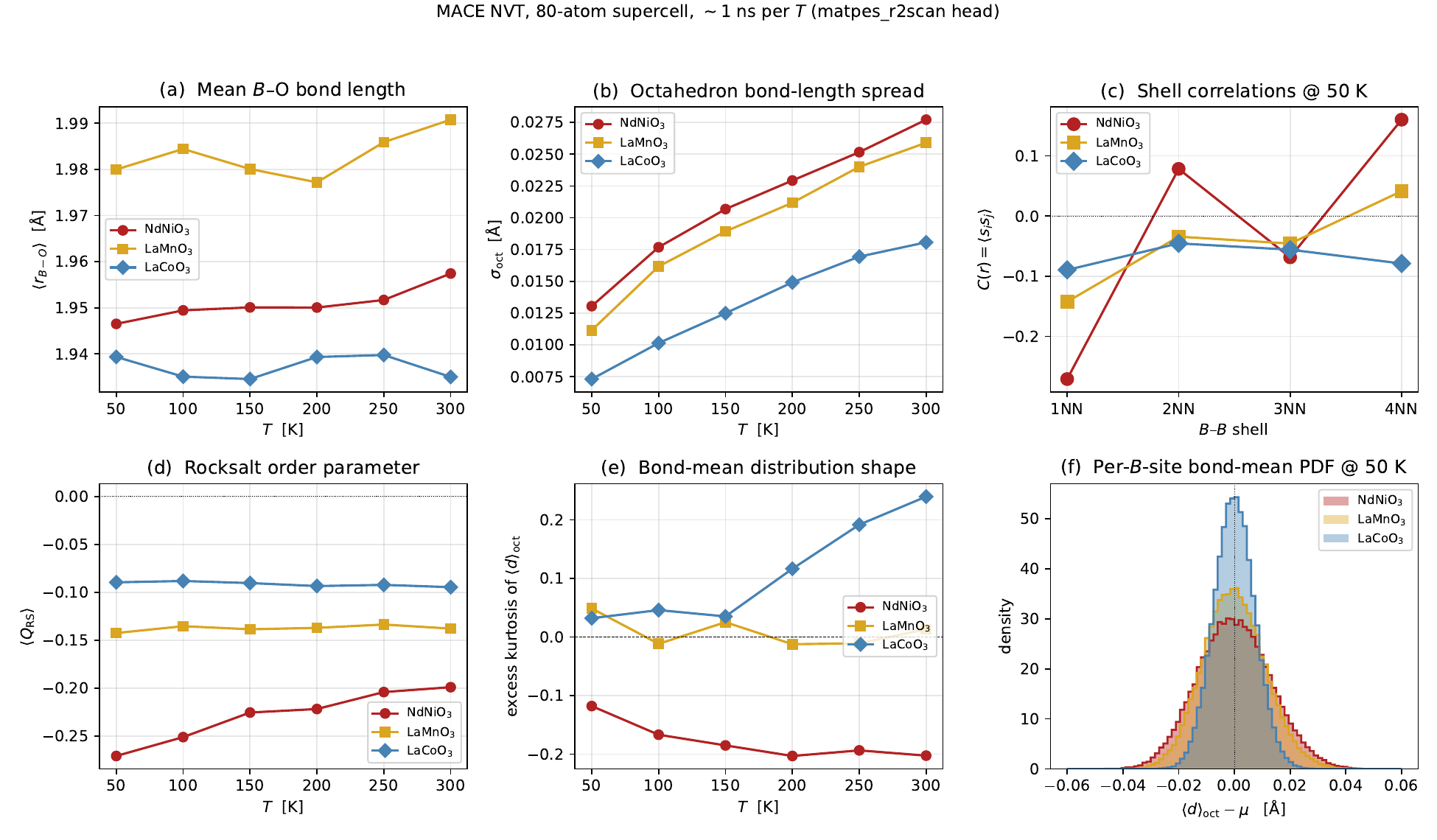}
\caption{\textbf{Figure 1 $|$ Structural signatures of \ce{NdNiO3} (firebrick),
\ce{LaMnO3} (goldenrod) and \ce{LaCoO3} (steelblue) across $T=50$--$300$\,K
under identical 80-atom MACE \texttt{matpes\_r2scan} NVT molecular
dynamics.}  (a)~Mean nearest-neighbour $B$--O bond
$\langle\overline{r}\rangle$ vs $T$: \ce{NdNiO3} and \ce{LaMnO3} expand by
$\sim\!+0.011$\,\Aa, \ce{LaCoO3} flat to $\pm 0.003$\,\Aa. (b)~Per-site
bond-length spread $\sigma_{\mathrm{oct}}$ grows linearly in all three;
\ce{NdNiO3} and \ce{LaMnO3} sit well above the \ce{LaCoO3} value.
(c)~Shell correlations $C(\mathrm{1NN/2NN/3NN/4NN})$
with $s_i=\mathrm{sign}(\overline{r}_i-\mathrm{median})$:
\ce{NdNiO3} shows the textbook rocksalt $(-,+,-,+)$ alternation at all $T$,
\ce{LaMnO3} and \ce{LaCoO3} are weakly negative with no alternation.
(d)~Rocksalt order parameter $Q_{\mathrm{RS}}$: \ce{NdNiO3} at $-0.27$,
\ce{LaMnO3} intermediate ($-0.14$), \ce{LaCoO3} weakest ($-0.09$).
(e)~Excess kurtosis $\gamma_{2}$ of the per-site bond-mean
distribution ($\gamma_{2}{=}0$ Gaussian, dashed): \ce{NdNiO3} sub-Gaussian
($-0.18$, two environments), \ce{LaMnO3} near-Gaussian, \ce{LaCoO3}
super-Gaussian and growing (single narrowing environment). (f)~Centred
per-site bond-mean histograms
at 50\,K with $\mu_{\mathrm{NNO}}{=}1.947$\,\AA, $\mu_{\mathrm{LMO}}{=}1.980$\,\AA,
$\mu_{\mathrm{LCO}}{=}1.939$\,\AA.  See Methods and Supplementary Section~S1 for descriptor
definitions.}
\label{fig:master}
\end{figure}

\begin{figure}[h]
\centering
\includegraphics[width=\textwidth]{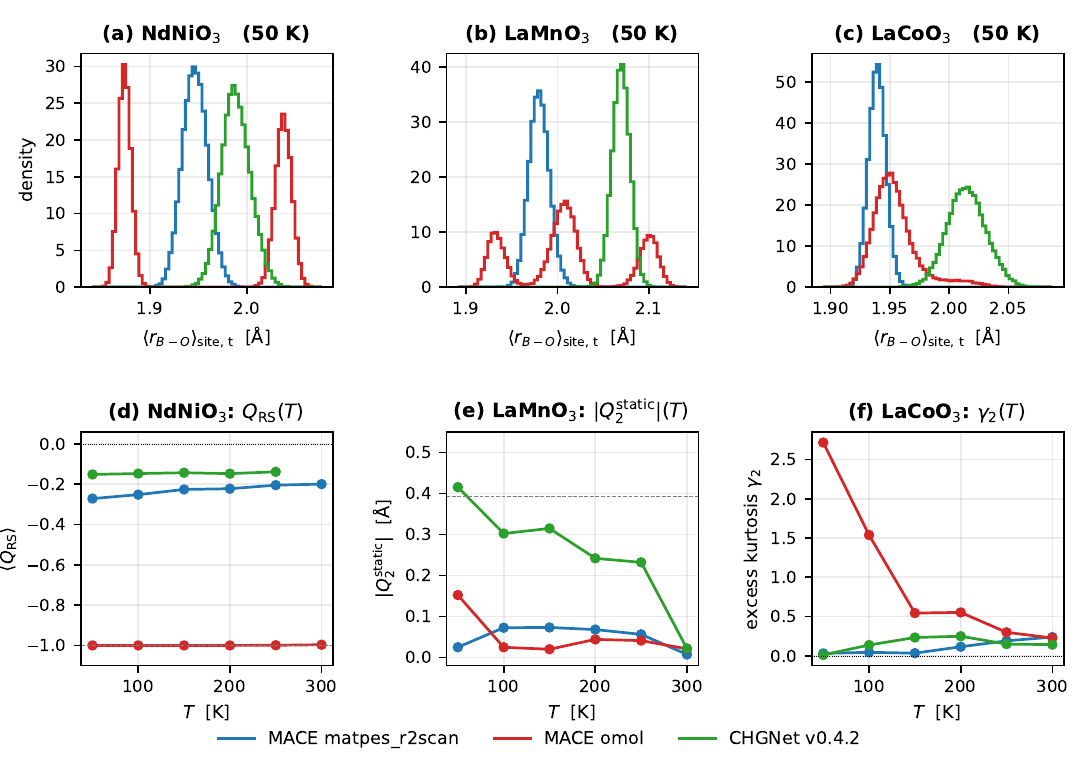}
\caption{\textbf{Figure 2 $|$ Three foundation potentials, three distinct
failure modes on identical 80-atom supercells.}
Colours: \texttt{matpes\_r2scan} (blue), \texttt{omol} (red),
CHGNet~v0.4.2 (green).
(a)--(c)~Per-site bond-mean PDFs at 50\,K.  By the Sarle bimodality
coefficient (BC; bimodal above $5/9$, ref.~\cite{Pfister2013}) only
\texttt{omol}--\ce{NdNiO3} is unambiguous (BC$=0.97$), matching the
breathing-mode rocksalt fingerprint; full BC table in Table~S2.
(d)~$Q_{\mathrm{RS}}(T)$ for \ce{NdNiO3}: only \texttt{omol} saturates
near the ideal $-1.0$ (dashed); \texttt{r2scan} reaches $-0.27$, CHGNet
stays at $-0.15$.
(e)~$|Q_{2}^{\rm static}|(T)$ for \ce{LaMnO3}: CHGNet alone produces a
static ferro-orbital JT of experimental
magnitude (dashed, $|Q_{2}|=0.39$\,\Aa{}~\cite{RodriguezCarvajal1998})
that thermally melts; \texttt{omol} shows a partial 50\,K distortion;
\texttt{r2scan} remains sub-0.025\,\Aa{}.
(f)~Excess kurtosis $\gamma_{2}(T)$ for \ce{LaCoO3}: \texttt{omol} is
mildly leptokurtic at low $T$; all three potentials agree on the absence of
any symmetry-breaking signature.}
\label{fig:cross_mlip}
\end{figure}

% ----------------------------------------------------------------------------
% Tables
% ----------------------------------------------------------------------------

\begin{table}
\centering
\small
\caption{\textbf{Single-head structural-footprint outcomes for the
production MACE \texttt{matpes\_r2scan} head.}  The single-head row
identifies which correlated-electron channel is captured by static
structural fingerprints in foundation MLIP molecular dynamics. ``Yes
(precursor)'' for \ce{NdNiO3} means the breathing-mode short-range
fluctuations are present in MD even though the static long-range
\textit{P}2$_{1}$/\textit{n} freezing is not; see
Table~\ref{tab:trilemma3} and Fig.~2 for the
cross-potential picture.}
\label{tab:trilemma}
\begin{tabular}{lccc}
\toprule
                              & \ce{LaCoO3}            & \ce{NdNiO3}                 & \ce{LaMnO3} \\
                              & ($d^{6}$)              & ($d^{7}$)                   & ($d^{4}$)   \\
\midrule
Ground space group            & \Rbarc                 & \Pbnm                       & \Pbnm \\
Electronic instability        & multiplet              & charge/ligand-hole          & JT + orbital order \\
Symmetry-breaking?            & no                     & yes                          & yes (orbital) \\
Lattice role                  & passive                & enabler / OP                 & OP itself \\
Static fingerprint?           & none                   & rocksalt long/short          & 1.91/1.97/2.18\,\Aa{} triplet \\
MACE captures?             & \textbf{no}            & \textbf{yes (precursor)}     & \textbf{no} \\
\bottomrule
\end{tabular}
\end{table}

\begin{table}
\centering
\scriptsize
\setlength{\tabcolsep}{4pt}
\caption{\textbf{Three foundation potentials $\times$ three correlated
oxides: cross-potential structural-fingerprint outcomes from the NVT $T$-series
(50--300\,K, identical 80-atom supercells, identical descriptor pipeline).}
Each entry lists the qualitative call together with the dominant single
observable; numerical ranges span the six $T$. ``$\approx$ expt''
refers to the structural channel only; it does not imply that the MD frame
is electronically insulating, nor that the potential predicts the correct
$T_{\mathrm{MI}}$.  \texttt{omol}--\ce{NdNiO3} is a kinetically trapped
breathing basin (flat $Q_{\mathrm{RS}}\!=\!-1.0$ across $T$), not a
thermodynamic phase prediction. $M_{\mathrm{FO}},M_{\mathrm{OO}}$ are
masked where $|Q_{2}^{\rm static}|\!<\!0.05$\,\Aa{} (sign of long axis
unresolved).}
\label{tab:trilemma3}
\begin{tabular}{lccc}
\toprule
                            & \texttt{matpes\_r2scan} & \texttt{omol}                 & CHGNet \\
\midrule
\textbf{NdNiO$_{3}$}        & short-range RS fluct.\ & frozen RS pattern ($\approx$ expt) & no RS pattern \\
expt: static breathing       & $Q_{\rm RS}=-0.20..-0.27$ & $Q_{\rm RS}=-1.000$ at all $T$ & $Q_{\rm RS}=-0.15$ \\
                             & $\sigma_{\rm oct}=0.013..0.028$\,\Aa & $\sigma_{\rm oct}=0.082$\,\Aa & $\sigma_{\rm oct}=0.015..0.033$\,\Aa \\
                             & BC$=0.35$              & BC$=0.81..0.97$ (bimodal)      & BC$=0.34$ \\
\midrule
\textbf{LaMnO$_{3}$}        & no static JT, Gaussian & partial-static FO @ 50\,K, dyn.\ above & static FO JT, melts with $T$ \\
expt: static C-type JT       & $|Q_{2}^{\rm static}|\!\le\!0.025$\,\Aa & $|Q_{2}^{\rm static}|=0.15\!\to\!0.02$\,\Aa & $|Q_{2}^{\rm static}|=0.42\!\to\!0.02$\,\Aa \\
                             & dyn.\ $Q_{\rm abs}=0.06..0.12$\,\Aa & dyn.\ $Q_{\rm abs}=0.17..0.19$\,\Aa & dyn.\ $Q_{\rm abs}=0.22..0.43$\,\Aa \\
                             & $M_{\rm FO},M_{\rm OO}\!\approx\!0$ (masked) & $M_{\rm FO}=0.67$ @ 50\,K & $M_{\rm FO}=1.0$, $M_{\rm OO}\!=\!0..0.25$ (FO, not C-type) \\
\midrule
\textbf{LaCoO$_{3}$}        & $\approx$ expt: Gaussian, no distortion & leptokurtic enhancement & small static ringing \\
expt: LS, undistorted        & $Q_{\rm RS}=-0.09$, BC$=0.33$ & $Q_{\rm RS}=-0.22..-0.29$, BC$=0.42..0.50$ & $Q_{\rm RS}=-0.09$, BC$=0.32$ \\
                             & $\gamma_{2}\!\sim\!0$  & $\gamma_{2}=+0.2..+2.7$         & $|Q_{2}^{\rm static}|=0.04..0.12$\,\Aa \\
\bottomrule
\end{tabular}
\end{table}

\end{document}